\documentclass[letter]{aa} 

\usepackage{epsfig}     
\usepackage{graphicx,color}     
\usepackage{amssymb}            
\usepackage{url}                
\usepackage{amsmath}            
\usepackage{rotating}                   
\usepackage{float}                      
\usepackage{textcomp}           
\usepackage{epstopdf}
\usepackage{dcolumn}
\usepackage{times}
\usepackage{tabularx}
\usepackage{hyperref}
\hypersetup{
    colorlinks,
    citecolor=blue,
    filecolor=blue,
    linkcolor=blue,
    urlcolor=blue,
    menucolor=black}
\usepackage{soul} 
\usepackage[english]{babel}
\usepackage{booktabs}
\usepackage{gensymb}
\usepackage{subfig}

\newcommand{\hrieuv}{HRI\textsubscript{EUV}\xspace}

\begin{document}


\title{Signatures of dynamic fibrils at the coronal base: Observations from Solar Orbiter/EUI}

\author{Sudip Mandal\inst{1},
Hardi Peter\inst{1},
Lakshmi Pradeep Chitta\inst{1}, 
Regina A. Cuadrado\inst{1}, 
Udo Sch\"{u}hle\inst{1},
Luca Teriaca\inst{1}, 
Sami K. Solanki\inst{1,2},
Louise Harra\inst{3,4}, 
David Berghmans\inst{5}, 
Fr\'{e}d\'{e}ric Auch\`{e}re\inst{6}, 
Susanna Parenti\inst{6}, 
Andrei N. Zhukov\inst{5,7},
\'{E}ric Buchlin\inst{6},
Cis Verbeeck\inst{5},
Emil Kraaikamp\inst{5},
Luciano Rodriguez\inst{5},
David M. Long\inst{8},
Conrad Schwanitz\inst{3,4},
Krzysztof Barczynski\inst{3,4},
Gabriel Pelouze\inst{6},
Philip J. Smith\inst{8},
Wei Liu\inst{9,10},
\and 
Mark C. Cheung\inst{11}
}

\institute{Max Planck Institute for Solar System Research, Justus-von-Liebig-Weg 3, 37077, G{\"o}ttingen, Germany \\
\email{smandal.solar@gmail.com}
\and
School of Space Research, Kyung Hee University, Yongin, Gyeonggi 446-701, Republic of Korea
\and
Physikalisch-Meteorologisches Observatorium Davos, World Radiation Center, 7260 Davos Dorf, Switzerland 
\and
ETH-Z\"{u}rich, Wolfgang-Pauli-Str. 27, 8093 Z\"{u}rich, Switzerland
\and
Solar-Terrestrial Centre of Excellence -- SIDC, Royal Observatory of Belgium, Ringlaan -3- Av. Circulaire, 1180 Brussels, Belgium
\and
Université Paris-Saclay, CNRS,  Institut d'Astrophysique Spatiale, 91405, Orsay, France
\and
Skobeltsyn Institute of Nuclear Physics, Moscow State University, 119992 Moscow, Russia
\and
UCL-Mullard Space Science Laboratory, Holmbury St. Mary, Dorking, Surrey RH5 6NT, UK
\and
Lockheed Martin Solar and Astrophysics Laboratory, Building 252, 3251 Hanover Street, Palo Alto, CA 94304, USA
\and
Bay Area Environmental Research Institute, NASA Research Park, Mailstop 18-4, Moffett Field, CA 94035, USA
\and
 CSIRO, Cnr Vimiera \& Pembroke Roads, Marsfield NSW, 2122, Australia
}

\abstract{
The solar chromosphere hosts a wide variety of transients, including dynamic fibrils (DFs) that are characterised as elongated, jet-like features seen in active regions, often through H$\alpha$ diagnostics. So far, these features have been difficult to identify in coronal images primarily due to their small size and the lower spatial resolution of the current EUV imagers. Here we present the first unambiguous signatures of DFs in coronal EUV data using high-resolution images from the Extreme Ultraviolet Imager (EUI) on board Solar Orbiter. Using the data acquired with the 174\,\AA\ High Resolution Imager (\hrieuv) of EUI, we find many bright dot-like features (of size 0.3-0.5 Mm) that move up and down (often repeatedly) in the core of an active region. In a space-time map, these features produce parabolic tracks akin to the chromospheric observations of DFs. Properties such as their speeds (14 km~s$^{-1}$), lifetime (332~s), deceleration (82 m~s$^{-2}$) and lengths (1293\,km) are also reminiscent of the chromospheric DFs. The EUI data strongly suggest that these EUV bright dots are basically the hot tips (of the cooler chromospheric DFs) that could not be identified unambiguously before because of a lack of spatial resolution. 
}

   \titlerunning{Signatures of dynamic fibrils at the coronal base}
   \authorrunning{Sudip Mandal et al.}
   \maketitle

\section{Introduction}\label{sec:intro}
Dynamic fibrils (DFs), the thin jet-like on-disc features, are typically observed through H$\alpha$ diagnostics of the solar chromosphere \citep{2007ASPC..368...27R}. They usually have a lifetime of 2 to 4 minutes and lengths between 1 to 4 Mm \citep{2005ApJ...624L..61D}. A combination of observations and numerical simulations strongly suggests that DFs are basically shock-driven phenomena \citep{2005ApJ...624L..61D,2006ApJ...647L..73H}. Magnetoacoustic waves that leak from the lower atmosphere are guided upward by magnetic fields and form shocks at chromospheric heights. These shocks then catapult chromospheric material upwards and produce jet-like DFs \citep{2007ApJ...666.1277H}. In a space-time (X-T) diagram, these DFs generate characteristic parabolic tracks that are a result of their up-and-down motion \citep{2006ApJ...647L..73H,2007ApJ...655..624D}. Interestingly, these properties are similar to those of other solar features such as type-I spicules (off-limb features) or quiet-Sun mottles (on-disc features) and, indeed, studies (\citealp[e.g.,][]{2007ApJ...660L.169R,2007ApJ...666.1277H}) have shown that all these different features are most probably a manifestation of a common underlying family of drivers related to shocks \citep{2004Natur.430..536D}.

Thus far, observations of DFs have mostly been restricted to the chromosphere (e.g., H$\alpha$ observations) and transition region (e.g., IRIS\footnote{Interface Region Imaging Spectograph; \cite{2014SoPh..289.2733D}.}
observations). Extreme Ultraviolet (EUV) observations of DFs using coronal imagers such as the Atmospheric Imaging Assembly \cite[AIA;][]{2012SoPh..275...17L} have particularly been hindered by their inadequate spatial resolution. For example, \citet{2016ApJ...817..124S} studied the evolution of DFs using co-ordinated SST\footnote{Swedish 1 m Solar Telescope; \cite{2003SPIE.4853..341S}.}, IRIS, and AIA data and found that the `bright rim' visible on top of a DF in IRIS 1400{\AA} channel images cannot be unambiguously identified in AIA channels due to their lower spatial resolution.

\begin{figure*}[!ht]
\centering
\includegraphics[width=0.98\textwidth,clip,trim=0cm 1cm 0cm 0cm]{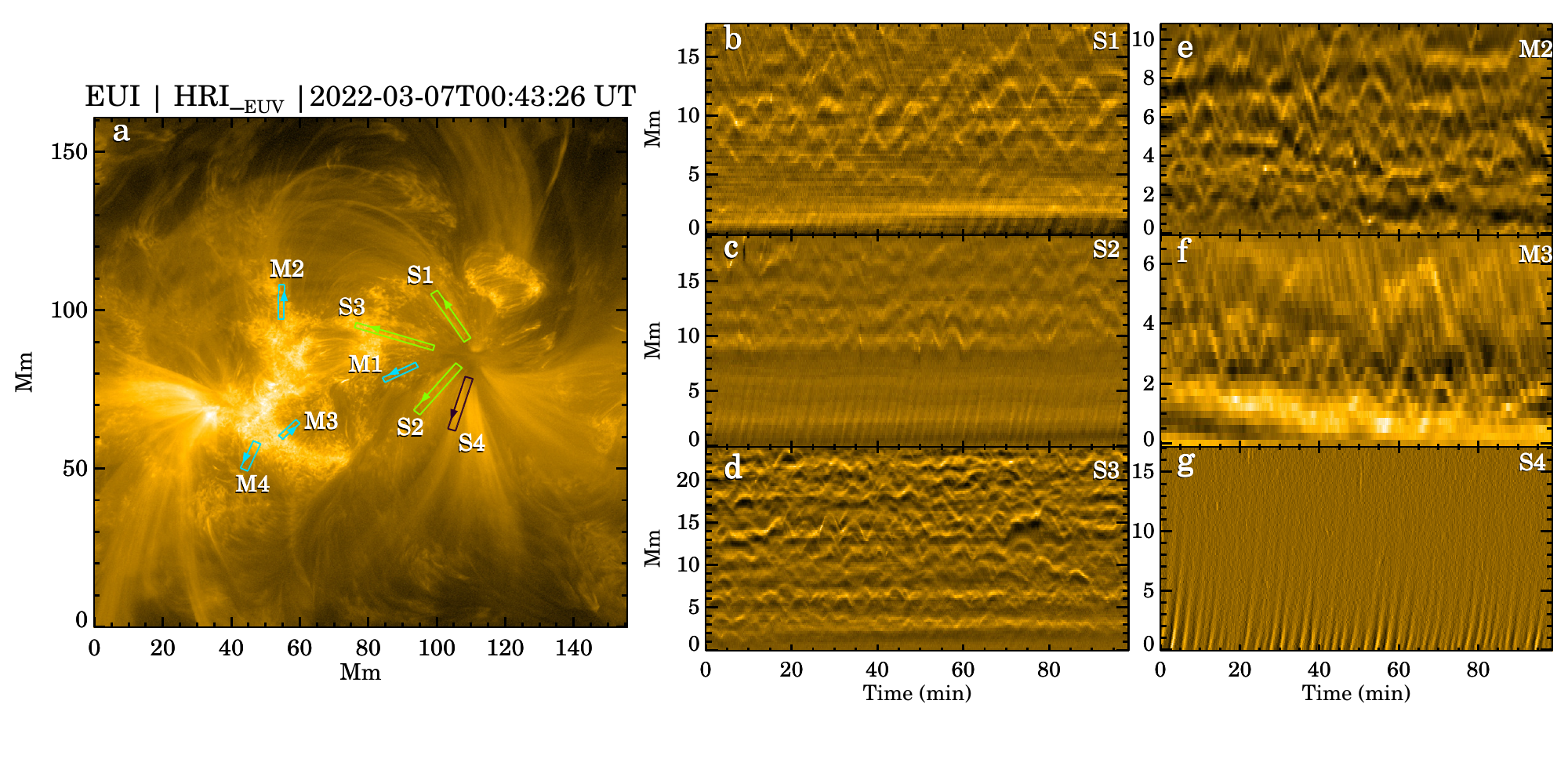}
\caption{Overview of the observed active region. Panel-a shows a sub-section of HRI$_{EUV}$ full field of view from 2022-03-07. The boxes in various colours represent the artificial slits that we used to derive space-time (X-T) maps, some of which are shown in Panels-b--g. These X-T maps are contrast enhanced by performing a boxcar smooth subtraction along the space axis. The arrows drawn on every slit point to the increasing y-axis of the corresponding X-T maps. An animated version of this figure is available \href{https://drive.google.com/file/d/1o_4jHA5JbyQtrpUBtB3ItE_s3HjF6ncc/view?usp=sharing}{here}.}
\label{fig:context}
\end{figure*}

In the present work, for the first time, we present unambiguous EUV signatures of such bright rims using high resolution, high cadence EUV observations from Solar Orbiter \citep{2020A&A...642A...1M}. We describe the data in Section 2, whereas Section 3 outlines the results. Finally we conclude by summarising our results in Section 4.

\section{Data}\label{sec:data}

We use EUV imaging data from the Extreme Ultraviolet Imager \cite[EUI;][]{2020A&A...642A...8R} on board Solar Orbiter. This particular dataset was taken by the 174~{\AA} High Resolution Imager (\hrieuv)
of EUI on 2022-03-07, between 00:35:06 and 02:14:02 UT, with a cadence of 5~s \citep[part of the SolO/EUI Data Release 5.0;][]{euidatarelease5}. At the time of this observation, Orbiter was located at 0.50\,AU distance from the Sun. This results in the EUI plate scale being 178 km per pixel on the Sun. To remove the effect of jitter from the EUI images, we employ a cross-correlation based image alignment technique similar to the one described in \citet{2022A&A...666L...2M}.

During this observation, Solar Orbiter was almost aligned with the Sun-Earth line (the angle spanned by Solar Orbiter, the Earth and the Sun being -0.58$\degree$). This allows us to use co-temporal and co-spatial data from the Solar Dynamics Observatory \cite[SDO;][]{2012SoPh..275....3P}. In particular, we use EUV images from the 171~{\AA} channel of the Atmospheric Imaging Assembly \cite[AIA;][]{2012SoPh..275...17L} and the line-of-sight (LOS) magnetograms from the Helioseismic and Magnetic Imager \cite[HMI;][]{2012SoPh..275..207S}, both onboard SDO. Although AIA 171~{\AA} and EUI 174~{\AA} passbands image plasma at comparable temperatures, the spatial resolution of AIA is significantly lower (plate scale of 0.6\arcsec or 435~km on the Sun). To support the EUI observations\footnote{This was coordinated as SOOP R-BOTH-HRES-HCAD-nanoflares \citep{2020A&A...642A...3Z,2022Perihelion}.}, AIA was running a special campaign at a 6~s (twice its nominal 12~s) cadence with half of its EUV channels (using only 131, 171, 193, and 304~{\AA}). This high-cadence sequence was interrupted every 96~s (i.e., 1 out of every 16 frames) in order to obtain synoptic, full-disk images.

The EUI and SDO data were aligned using a combination of FITS keywords and visual inspection. This is sufficient for the current study, because we only use the AIA data for qualitative comparisons. Lastly, while comparing the AIA and EUI data, we take into account the difference in light travel times between Sun-EUI and Sun-AIA. All the times that we quote in the paper are times at 1\,AU.

\section{Results}\label{sect:results}

\begin{figure}
\centering
\includegraphics[width=0.45\textwidth,clip,trim=0cm 0.3cm 0cm 0.5cm]{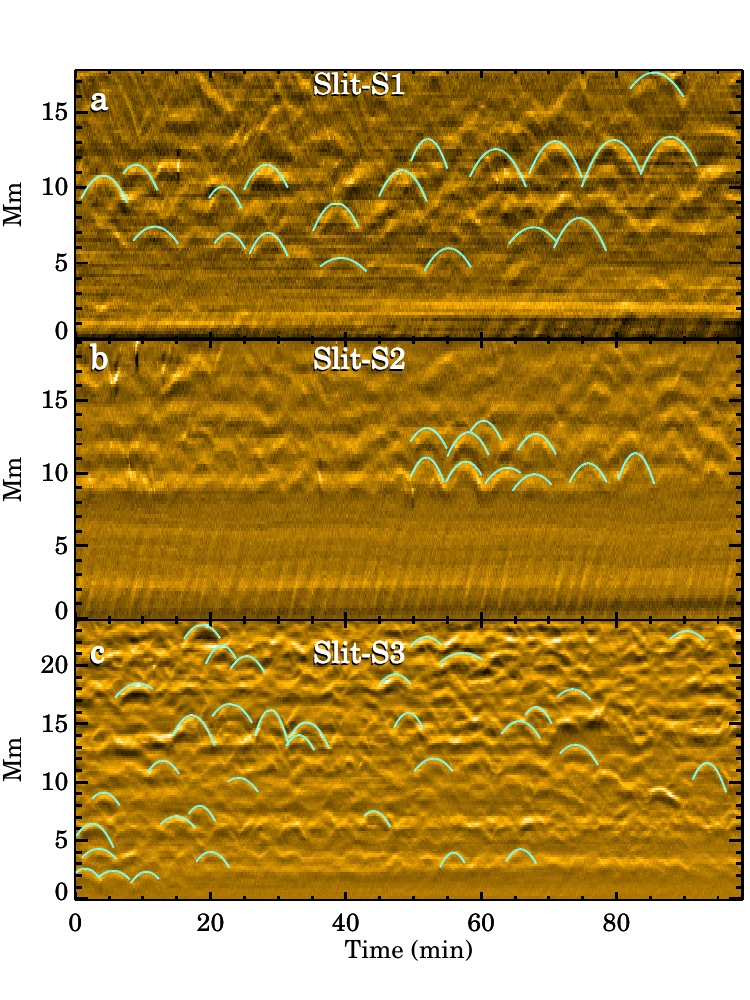}
\caption{Parabolic trajectories of dynamic fibrils. Panels a,b,c show the contrast enhanced X-T maps from slit-S1,S2,S3, respectively. In each of these panels, the curves in cyan outline the fitted parabolas to the visually identified bright tracks. See Sect.~\ref{sect:xt} for details.}
\label{fig:xt}
\end{figure}
\begin{figure*}
\centering
\includegraphics[width=0.95\textwidth,clip,trim=0cm 0cm 1cm 0cm]{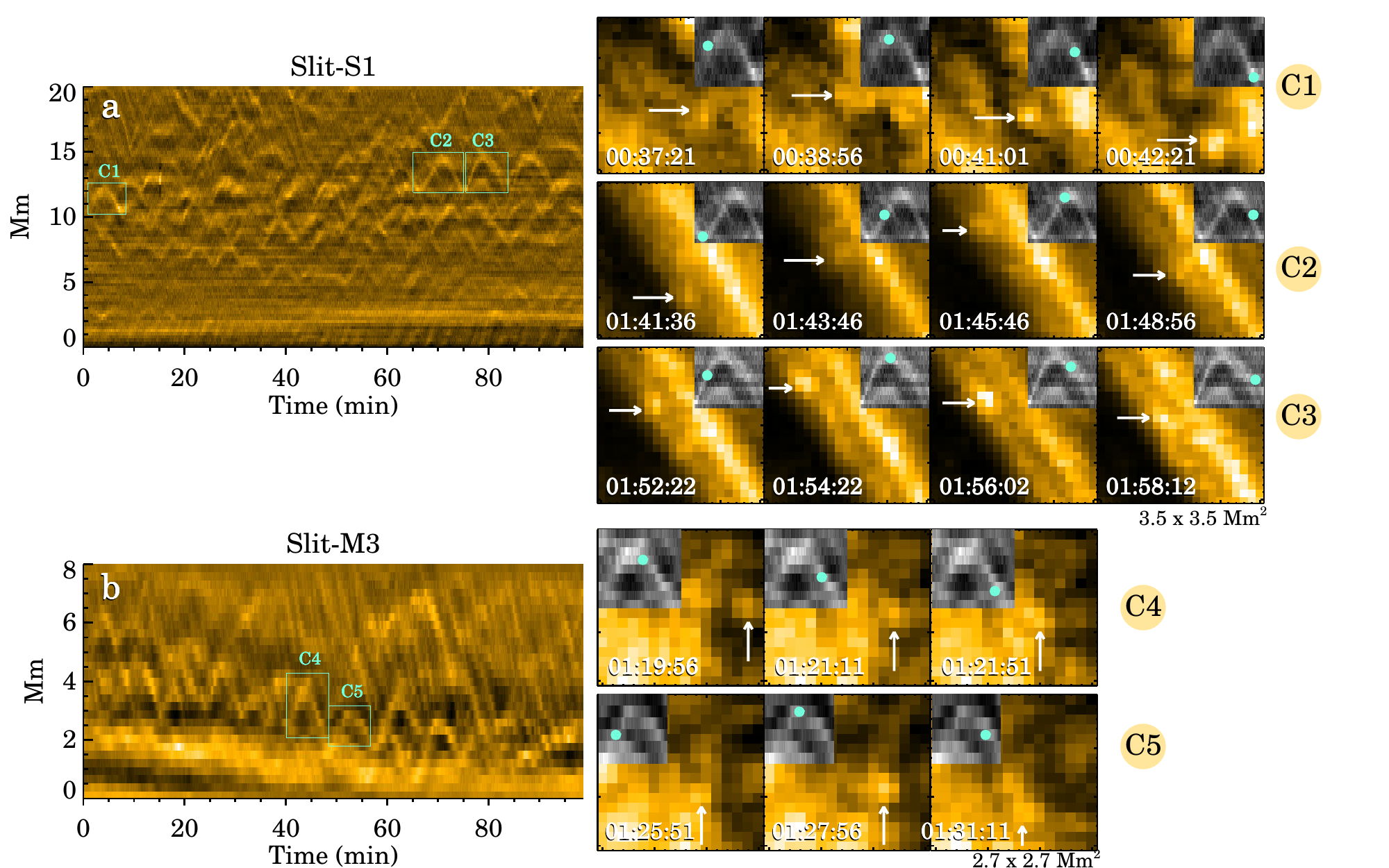}
\caption{A closer look at up and down motions. Panel-a shows the contrast enhanced X-T map from Slit-S1. The cyan boxes C1, C2 and C3 highlight the chosen tracks, whose snapshots are presented in the adjacent panels. In each of these panels, the brightening that creates the track is highlighted by an arrow wherein the track itself (from the X-T map) is shown in the greyscale inset panel. Cyan dots overlaid on each of these inset tracks are indicative of the time at which the EUI snapshots were taken. Panel-b shows the contrast enhanced X-T map from Slit-M3 along with two cases highlighted as C4 and C5.}
\label{fig:indi}
\end{figure*}

In order to identify the DFs and to quantify their properties, we locate them in the EUI movie and investigate their evolution using space-time diagrams. Figure~\ref{fig:context}a presents a snapshot from the EUI dataset which primarily encompasses the active region NOAA12960. In and around this active region (AR), we find typical coronal structures such as closed loops, moss regions, fan loops, and jets. A closer look at the event movie (available online)
reveals the ubiquitous presence of tiny bright blob-like or front-like structures that move back and forth with time. These are the features that we refer to as the EUV signatures of DFs\footnote{We do not really observed the elongated fibrils in EUV, but rather only tips of them.} and are the subject of this study.

\subsection{Capturing the EUV dynamic fibrils}\label{sect:xt}

In order to capture the observed motions of these DFs, we generate multiple space-time (X-T) maps by placing artificial slits at several locations as shown in Fig.~\ref{fig:context}a. Each slit is 14 pixels wide and the final X-T maps are generated after averaging over these 14 pixels. Moreover, to highlight the bright tracks, we perform a boxcar (of 20 pixels) smooth subtraction along the transverse direction (i.e., along y-axis) in each of these X-T maps.

In this overview figure (Fig.~\ref{fig:context}), names of slits that are closer to the leading sunspot of the AR start with an $\rm{S}$ (for spot) and the ones that are away from it, overlying the moss regions, start with an $\rm{M}$ (for moss). Through this slit arrangement, we aim to cover a variety of magnetic environments. For example, slit-$\rm{S1}$ lies between two closed loops, slit-$\rm{S4}$ lies along a fan loop, slit-$\rm{M1}$ overlies a moss region close to the spot, whereas slits $\rm{M2}$, $\rm{M3}$ and $\rm{M4}$ are also on moss regions but away from the leading spot in the AR. Panels-b to g in Fig~\ref{fig:context} present six out of the eight X-T maps that we have derived.

Already at a first glance, all of these X-T maps (except panel-g), contain bright tracks that appear to have parabolic shapes. Furthermore, many of these tracks are also repetitive. As mentioned earlier, these are basically the signatures of dynamic fibrils that move up and down along the slits. Before we go on to examine these parabolic tracks in detail, we briefly discuss the slit-$\rm{S4}$ X-T map (panel-g). This map does not contain any parabolic tracks but rather only shows straight, periodic slanted ridges. These straight ridges are manifestations of the slow magneto-acoustic waves that propagate along the magnetic field \cite[e.g.,][]{2012A&A...546A..50K}. In fact, such slanted ridges can also be seen in the bottom parts of the $\rm{S1}$ (Fig.~\ref{fig:context}b) and $\rm{S2}$ (Fig.~\ref{fig:context}c) maps. 

In each X-T map, we fit the individual seemingly parabolic bright tracks with a parabola and derive parameters such as deceleration, maximum speed (during ascent or descent) as well as maximum length (also referred as `height' in the literature) and lifetime. The fitted parabolas are overplotted in cyan in Fig.~\ref{fig:xt}. In total 98 profiles are fitted across all X-T maps. Although most of the tracks can be well approximated by a parabola, there are a few cases, e.g., between T=(45,50)\,min and X=(15,17)\,Mm in Fig.~\ref{fig:xt}c, where the paths appear closer to a triangle (i.e., with constant velocity). While fitting the parabolic tracks, we only chose the ones for which the signal is unambiguous (i.e., tracks that are relatively free from overlaps with other DFs). To detect a bright track in the first place, we use Gaussian fits along the transverse direction of an X-T map. Once a bright track is identified (through the centers of the fitted Gaussians), we then fit a parabola to that detected track and determine its parameters. Given that most tracks are inevitably partially visible i.e., with a missing part during the ascending or descending phase, we first extrapolated the fitted curve to make it symmetric and then calculated its lifetime.

\begin{figure}
\centering
\includegraphics[width=\columnwidth,clip,trim=0cm 0cm 1cm 0cm]{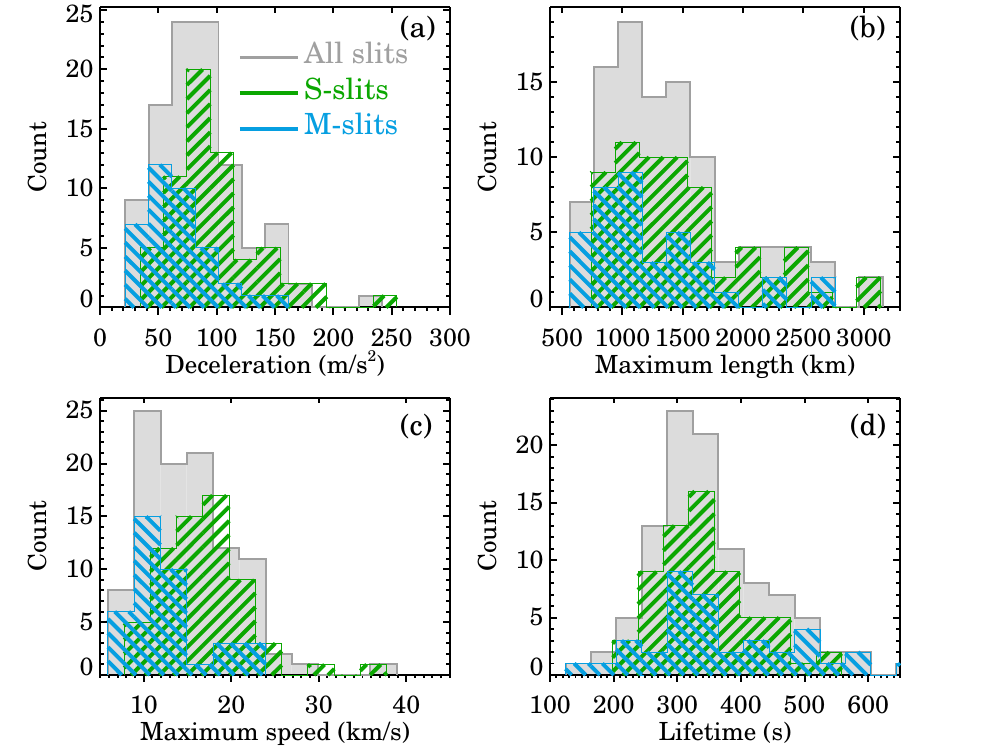}
\caption{Properties of DFs in EUV. Distributions of fitted parameters such as deceleration (panel-a), maximum length (panel-b), maximum speed (panel-c) and lifetime (panel-d). In each panel, the grey shaded histogram presents the overall distribution of all analysed X-T map profiles (sample size $n=98$) wherein the distributions derived only for $\rm{S}$- and $\rm{M}$ -slits ($n=61$ and $37$, respectively) are shown in green and cyan colours, respectively. Further details are listed in Table~\ref{tab:table}. }
\label{fig:histogram}
\end{figure}

\subsection{Inspection of selected individual EUV dynamic fibrils}

As mentioned earlier, the event movie reveals multiple bright structures that move up and down with time. To give some examples, through a series of snapshots in Fig.~\ref{fig:indi}, we highlight how a blob-like moving bright feature creates a parabolic track in the X-T map. Figure~\ref{fig:indi}a presents the X-T map from slit-$\rm{S1}$ which we recall, is located in-between two closed loops (see Fig.~\ref{fig:context}a). In all three highlighted cases (C1, C2, C3), we find a small brightening of size $\approx$ 0.5 Mm (i.e., only some 2 to 3 EUI pixels) exhibiting up and down motions (see the snapshots in the panels on the right of the figure). The cases C2 and C3 further highlight the repetitive nature of these phenomena. The other interesting aspect to note here is that of the change in size of these features as they evolve with time. For example, in C1 we find that the size increases with time whereas in C2 it remains almost constant and in C3, it initially increases but then remains constant for the rest of the time.

In another example, Fig.~\ref{fig:indi}b shows the X-T map from $\rm{M3}$ which is located on top of a moss region (see Fig.~\ref{fig:context}a). Interestingly, in the two highlighted cases from this map i.e., C4 and C5, we find the brightenings to be of even smaller sizes ($\approxeq$0.3 Mm) compared to those in Fig.~\ref{fig:indi}a. Through these examples we therefore confirm that the parabolic tracks in the X-T maps are indeed due to movements of blob-like bright EUV features.

\begin{table}
    \centering
    \caption{Statistical properties of EUV dynamic fibrils}
    \resizebox{\columnwidth}{!}{%
    \begin{tabular}{l|rcccc}
    \hline\hline
         \multicolumn{2}{@{}c@{}}{} &Deceleration$^{\dagger}$ & Max. length$^{\dagger}$ & Max. Speed$^{\dagger}$ & Lifetime \\
         \multicolumn{2}{@{}c@{}}{}  & [m s$^{-2}$] & [km] & [km s$^{-1}$] & [s] \\
    \hline
         \multicolumn{2}{@{}c@{}}{All slits}& 82 & 1293 & 14.0 & 332 \\
         \multicolumn{2}{@{}c@{}}{S-slits} & 90 & 1365 & 16.6 & 332\\
         \multicolumn{2}{@{}c@{}}{M-slits} & 66 & 1100 & 11.3 & 336\\
    \hline
        K-S & $D$ & 0.47 & 0.31 & 0.52 & 0.20 \\
        test\tablefootmark{a} & $p$ & 3.57$\times$10$^{-5}$ & 0.02 & 2.58$\times$10$^{-6}$ & 0.27\\
    \hline
    \end{tabular}
    }
\tablefoot{ \tablefoottext{$\dagger$} {These are projected values and thus, lower limits to their true values. \tablefoottext{a}{Kolmogorov–Smirnov (K-S) test comparing samples from the S-slits and M-slits}.}
}
  \label{tab:table}
\end{table}

\begin{figure*}[!htb]
\centering
\includegraphics[width=0.97\textwidth,clip,trim=0cm 0cm 0cm 0cm]{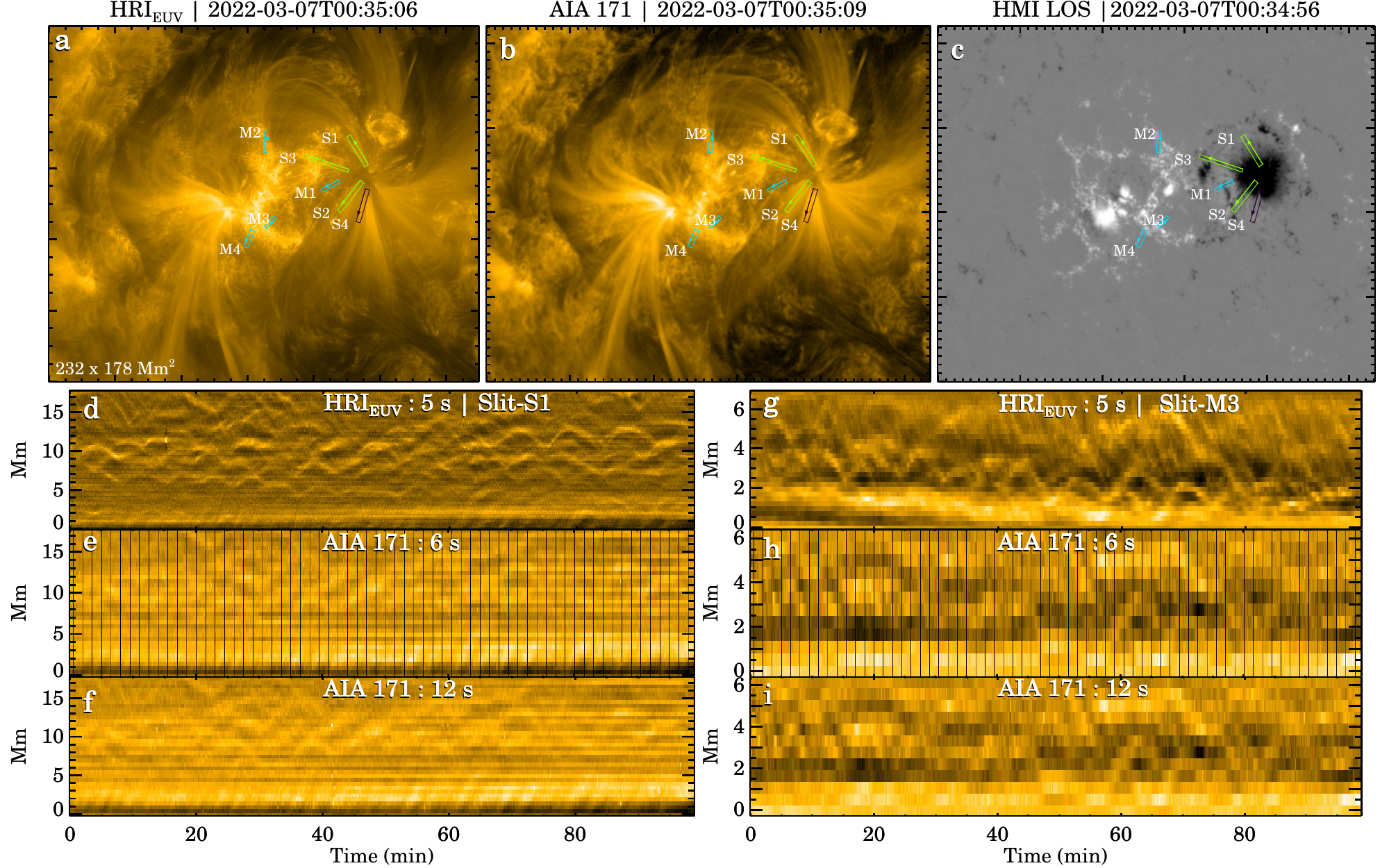}
\caption{Comparison between the EUI and AIA views of DFs. Panel-a presents a snapshot from the EUI 174~{\AA} timeseries, while a co-aligned, co-temporal AIA 171~{\AA} image is shown in panel-b and the HMI line of sight (LOS) magnetogram is displayed in panel-c. All these three panels have the slits from Fig. 1a overplotted on them. Panels-d, e and f present the slit-S1 X-T maps (contrast enhanced) derived using the EUI data, the AIA 171~{\AA} 6~s image sequence and the AIA 171~{\AA} 12~s sequence, respectively. The same but for slit-M3 are shown in panels g, h and i. The vertical lines in panels-e and h represent the missing frames in that AIA dataset. }
\label{fig:aia_eui_comp}
\end{figure*}

\subsection{Statistical properties of EUV dynamic fibrils}

In this section we examine the statistical properties of these parabolic tracks. Of course parameters such as speeds and lengths suffer from projection effects and thus, the values that are quoted here are apparent ones. The results of this statistical analysis are summarised in Fig.~\ref{fig:histogram}. Different panels shown there outline the overall distributions (in grey) of deceleration (panel-a), maximum length (panel-b), maximum speed (panel-c) and lifetime (panel-d), respectively. We further segregate each of these parameters according to the slit type where the respective DFs are found. The histograms for the respective subsets of the $\rm{S}$-slits (green) and $\rm{M}$-slits (cyan) are overplotted in Fig.~\ref{fig:histogram}. We find that the median values of deceleration, maximum length and maximum speed distributions of the DFs for $\rm{S}$-slits are higher than that for $\rm{M}$-slits (see Table~\ref{tab:table}). Lifetime distributions, however, have similar medians for both slit classes.

To check the statistical significance of the differences between the M-slits in moss regions and the S-slits close to sunspots, we perform a two sided Kolmogorov–Smirnov (K–S) test between the samples from the two slit types. Basically, the K-S test compares the empirical cumulative distribution functions (ECDFs) of two samples and decides whether those two samples are drawn from the same parent population or not \citep{ber2014}. The test statistic $D$ is a measure of the maximum distance between two ECDFs, while the probability $p$ endorses the null hypothesis that the two samples come from the same underlying population. As often assumed, we reject the null hypothesis if $p \leq 0.05$ and conclude that the two samples are inherently different from each other. K-S test results of our samples, as summarised in the last rows of Table~\ref{tab:table}, validate the statistical significance of our earlier observation about the differences in the DF properties between $\rm{S}$-slits and $\rm{M}$-slits. 

Lastly, we perform a brief comparison between the overall distributions of DFs that we have obtained in this study to that of the chromospheric observation of DFs by \citet{2007ApJ...655..624D}. By comparing the median values, we find that our deceleration median of 82 m~s$^{-2}$ is significantly lower than the value of 136 m~s$^{-2}$ given by \citet{2007ApJ...655..624D} (note that the deceleration distribution shown in their paper is also quite broad). In case of lifetime, we obtain a higher median of 332~s as opposed to 250~s found by \citet{2007ApJ...655..624D}. Interestingly, the other two median values, i.e., of maximum length (1293 km) and maximum speed (14 km~s$^{-1}$), are comparable to the ones found by them. Further comparisons with the results by \citet{2007ApJ...655..624D} are presented in Appendix~\ref{app1}.

\subsection{SDO view of EUV dynamic fibrils}

Given the favourable (very small) angle between the SDO and Solar Orbiter, we also analyse the same region as seen by EUI, now using data from AIA, employing the same techniques as for the EUI data. Here we also add the magnetic field information acquired by HMI. Along with the EUI image, Fig.~\ref{fig:aia_eui_comp} presents co-temporal images from the co-aligned AIA 171~{\AA} data (panel-b) and the HMI line-of-sight magnetogram (panel-c). 
As expected, the features in the AIA 171~{\AA} image appear to be quite similar to those of EUI 174~{\AA} image while the magnetogram reveals a large-scale bipolar-type magnetic configuration of the host AR (NOAA12960). Furthermore, the overlaid slits on the magnetogram re-confirms our slit classification, i.e., the $\rm{S}$-slits touch the leading spot, whereas the $\rm{M}$-slits are placed away from it in plage-type areas that are expected below moss regions. The main reason to include SDO data here is to compare the visibility of the parabolic tracks of the DFs between EUI and AIA in the corresponding X-T maps. For illustration, we chose two of the slits we already investigated with EUI, i.e., slit-$\rm{S1}$ which is located close to the spot and slit-$\rm{M3}$ which is located far away from that spot and lies on-top of a moss region (cf. Fig.~\ref{fig:aia_eui_comp}a-c). The AIA 6~s cadence dataset has regular data gaps (see Sec.~\ref{sec:data}),
which somewhat distract our attention in those corresponding X-T maps (Fig~\ref{fig:aia_eui_comp}e,h) and therefore, we also used the 12~s cadence (the usual AIA cadence) data and regenerate those X-T maps (Fig.~\ref{fig:aia_eui_comp}f,i).

By comparing the X-T maps of these slits between EUI and AIA (Fig.~\ref{fig:aia_eui_comp}d,e,f and f,g,i) we find two main results: (1) there are significantly fewer parabolic tracks visible in AIA as compared to EUI and, (2) some of the AIA tracks are ambiguous and can only be identified in hindsight, i.e., only after seeing the corresponding EUI map. For example, the ambiguous signal in the AIA map (Fig~\ref{fig:aia_eui_comp}e) around T=(65,85)\,min and x=(10,13)\,Mm is clearly distinguishable in the corresponding EUI map (Fig~\ref{fig:aia_eui_comp}d). Considering the difference of a factor of ca. 2.5 in their spatial resolutions (cf.\ Sect.~\ref{sec:data}), it is not surprising that the DFs are much harder to detect, or in many cases not visible at all in AIA image sequences. The scenario that these small-scale features are not always detectable in AIA images is somewhat similar to the visibility of campfires in simultaneous EUI and AIA images as noted by \citet{2021A&A...656L...4B} and \citet{2021A&A...656L..16M}. This could also explain why \citet{2016ApJ...817..124S} were not able to unambiguously detect counterparts of chromospheric DFs in AIA coronal data.


\section{Discussion and summary}
Using high resolution, high cadence EUV images from EUI 174~{\AA}, we find bright blob-like features that repeatedly move up and down and leave parabolic tracks in X-T maps. The sizes of these blobs are between 0.3 Mm to 0.5 Mm. Other properties such as their speeds or lifetimes are very similar to chromospheric dynamic fibrils observed in H$\alpha$ by \citep{2007ApJ...655..624D}. Hence, we conclude that these features are the EUV signatures of dynamic fibrils (DFs). As mentioned in the introduction, DFs are shock driven chromospheric phenomena \citep{2006ApJ...647L..73H} and hence, the brightenings that we find in EUI 174~{\AA} images, are probably the hot tips of those fibrils. Furthermore, appearance of these bright parabolic tracks is strikingly similar to that of the `bright rims' in IRIS 1400~{\AA} images as noted by \citet{2016ApJ...817..124S}. Unfortunately, the transition region images from the coordinated IRIS observation, have a negligible overlap with EUI, both in terms of spatial and temporal coverage and thus, we lack the lower atmospheric view of the EUV DF features we observed with EUI. Moreover, in the absence of such coordinated spectroscopic observations, we are unable to conclusively determine whether the EUV emission that we see in EUI, comes from a coronal plasma ($\log$ T= 6) or from more of a slightly cooler transition region material ($\log$ T= 5.4). Future observations are needed to examine this further.

One of the interesting results we obtained from this study relates the properties of these DFs to their location of origin. For example, DFs that originate closer to a spot (e.g., $\rm{S}$-slits) are faster, longer and decelerate more strongly than those that emerge from a moss region (e.g., $\rm{M}$-slits). Using chromospheric observations, \citet{2007ApJ...655..624D} also found significant differences between DFs that come from a dense plage region (having predominantly vertical magnetic field) and those that originate from a less dense plage region (with more inclined field). These authors explained the observed differences in terms of a combination of shock waves and the local magnetic field topology. For instance, a vertical magnetic field guides 3-minute oscillations, that possess less power compared to 5-minute oscillations that are channelled along a more inclined field. As a result, DFs in inclined field regions have higher speeds (owing to higher driving power; \citealp{2013ApJ...776...56R}) and lower deceleration (because only a component of gravity acts along the field lines; \citealp{2006ApJ...647L..73H}). 
However, there is more to the story. As DFs are shock driven phenomena, we expect a stronger shock (such as in inclined fields) to produce a greater deceleration \citep{2007ApJ...655..624D}. In our study, we indeed find such signatures. For example, $\rm{S}$-slit DFs show higher speeds than $\rm{M}$-slits and at the same time, these $\rm{S}$-slit DFs also suffer greater deceleration than $\rm{M}$-slits. This is also evident through the correlation analysis that we present in Appendix~\ref{app1}. This further strengthens our explanation of the observed bright tracks as signatures of DFs.

Future studies using EUI data that accommodate DFs from a wide variety of magnetic regions will be helpful in further understanding such regional dependencies. A central element in such studies should be the combination of simultaneous observations of transition region and chromospheric plasma, requiring careful coordination between Earth-based facilities and Solar Orbiter. This will help in our understanding of the role of DFs in shaping the upper solar atmosphere in ARs.  

To conclude, by analysing high resolution high cadence EUV images from Solar Orbiter, we find bright dot-like features that move up and down with time (often repeatedly) and produce parabolic tracks in X-T diagrams. Their properties, such as speeds, lifetime, lengths, and deceleration, strongly suggest that these bright dots in the EUV images are a signature of the hot gas at or close to the tips of the chromospheric dynamic fibrils (DFs).

\begin{acknowledgements}
We thank the anonymous reviewer for the encouraging comments and helpful suggestions.. Solar Orbiter is a space mission of international collaboration between ESA and NASA, operated by ESA. The EUI instrument was built by CSL, IAS, MPS, MSSL/UCL, PMOD/WRC, ROB, LCF/IO with funding from the Belgian Federal Science Policy Office (BELSPO); the Centre National d’Etudes Spatiales (CNES); the UK Space Agency (UKSA); the Bundesministerium f\"{u}r Wirtschaft und Energie (BMWi) through the Deutsches Zentrum f\"{u}r Luft- und Raumfahrt (DLR); and the Swiss Space Office (SSO). L.P.C. gratefully acknowledges funding by the European Union. Views and opinions expressed are however those of the author(s) only and do not necessarily reflect those of the European Union or the European Research Council (grant agreement No 101039844). Neither the European Union nor the granting authority can be held responsible for them. The ROB team thanks the Belgian Federal Science Policy Office (BELSPO) for the provision of financial support in the framework of the PRODEX Programme of the European Space Agency (ESA) under contract numbers 4000134474, 4000134088, and 4000136424. WL and MCMC acknowledge support from NASA's SDO/AIA contract (NNG04EA00C) to LMSAL. AIA is an instrument on board the Solar Dynamics Observatory, a mission for NASA's Living With a Star program. WL is also supported by NASA grants 80NSSC21K1687 and 80NSSC22K0527. D.M.L. is grateful to the Science Technology and Facilities Council for the award of an Ernest Rutherford Fellowship (ST/R003246/1). S.P. acknowledges the funding by CNES through the MEDOC data and operations center. This research has made use of NASA’s Astrophysics Data System. The authors would also like to acknowledge the Joint Science Operations Center (JSOC) for providing the AIA data download links.
\end{acknowledgements}

\bibliographystyle{aa}
\bibliography{ref_DF}

\begin{appendix}

\section{Correlations and further comparisons}\label{app1}

\begin{figure}
\centering
\includegraphics[width=\columnwidth,clip,trim=0cm 0cm 0.1cm 0cm]{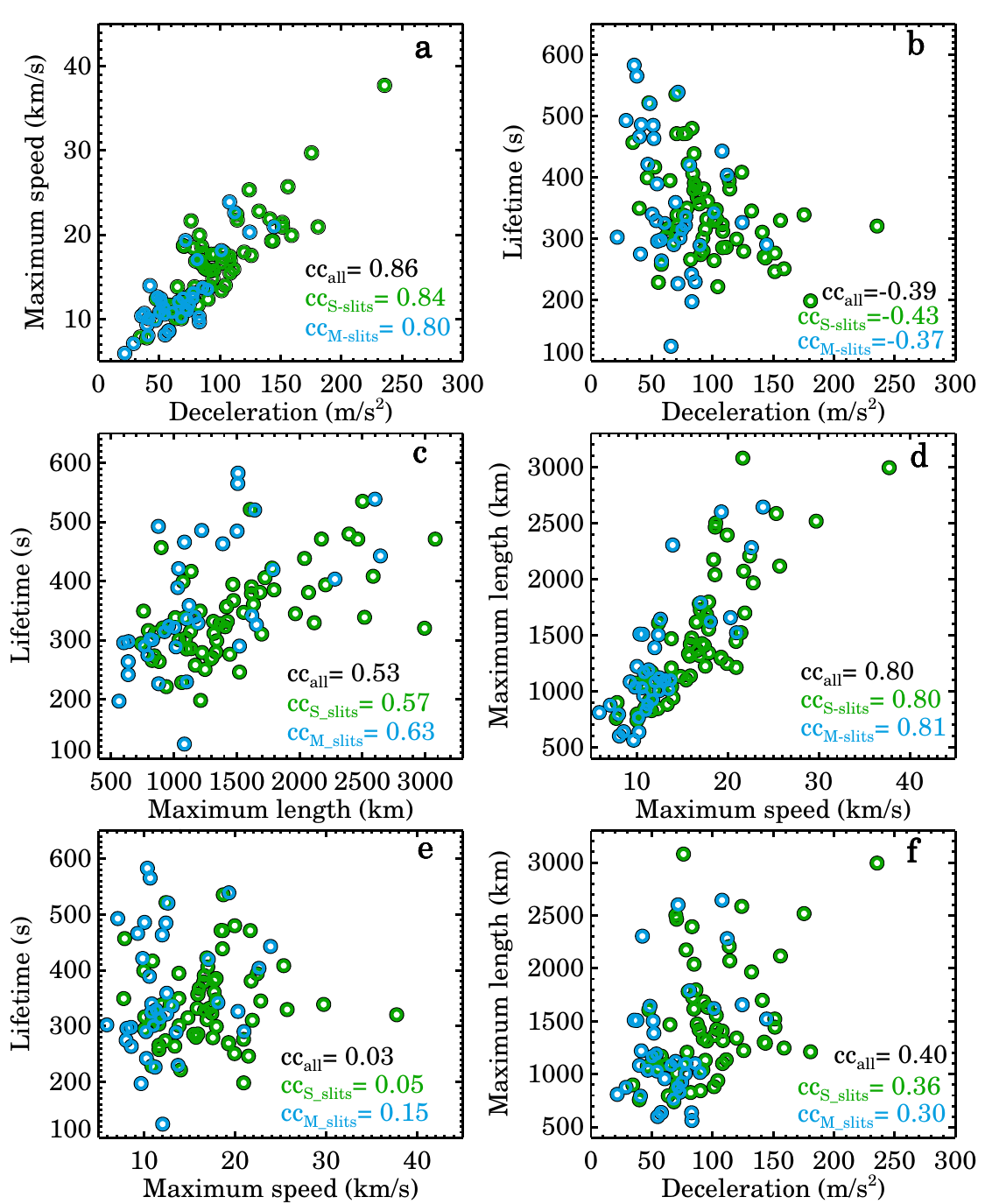}
\caption{Correlation analysis between different parameters of the fitted parabolas of EUV dynamic fibrils. Derived correlation coefficients (cc) for all three cases i.e., for all slits, only $\rm{S}$-slits and only $\rm{M}$-slits, are also printed on each panel. The green and blue circles represent DFs from the S-slits and M-slits, respectively. }
\label{fig:correlation}
\end{figure}

We analyse here the inter-relationships between various DF parameters by examining their scatter plots. From Fig.~\ref{fig:correlation}a, we find a strong positive correlation between deceleration and maximum velocity. This means that a faster DF suffers more deceleration, while a DF that lives longer is subjected to less deceleration as seen from the negative correlation in Fig.~\ref{fig:correlation}b. In Fig.~\ref{fig:correlation}d, we find that DFs that have higher speeds also have greater lengths. A positive correlation is further found between maximum length and lifetime (Fig.~\ref{fig:correlation}c), although with a significant spread for larger lengths. Lastly, no significant correlation is found either between maximum velocity and lifetime (Fig.~\ref{fig:correlation}e) or among deceleration and maximum length (Fig.~\ref{fig:correlation}f). 
All of these trends do match well with the findings of \citet{2007ApJ...655..624D} who also performed a similar analysis but for chromospheric observations of DFs. As explained in \citet{2007ApJ...655..624D}, these correlations are basically a manifestation of the fact that DFs are a shock driven phenomenon.

\section{Multi-wavelength view}
\begin{figure*}[!htb]
\centering
\includegraphics[width=\textwidth,clip,trim=0cm 0.3cm 0cm 0cm]{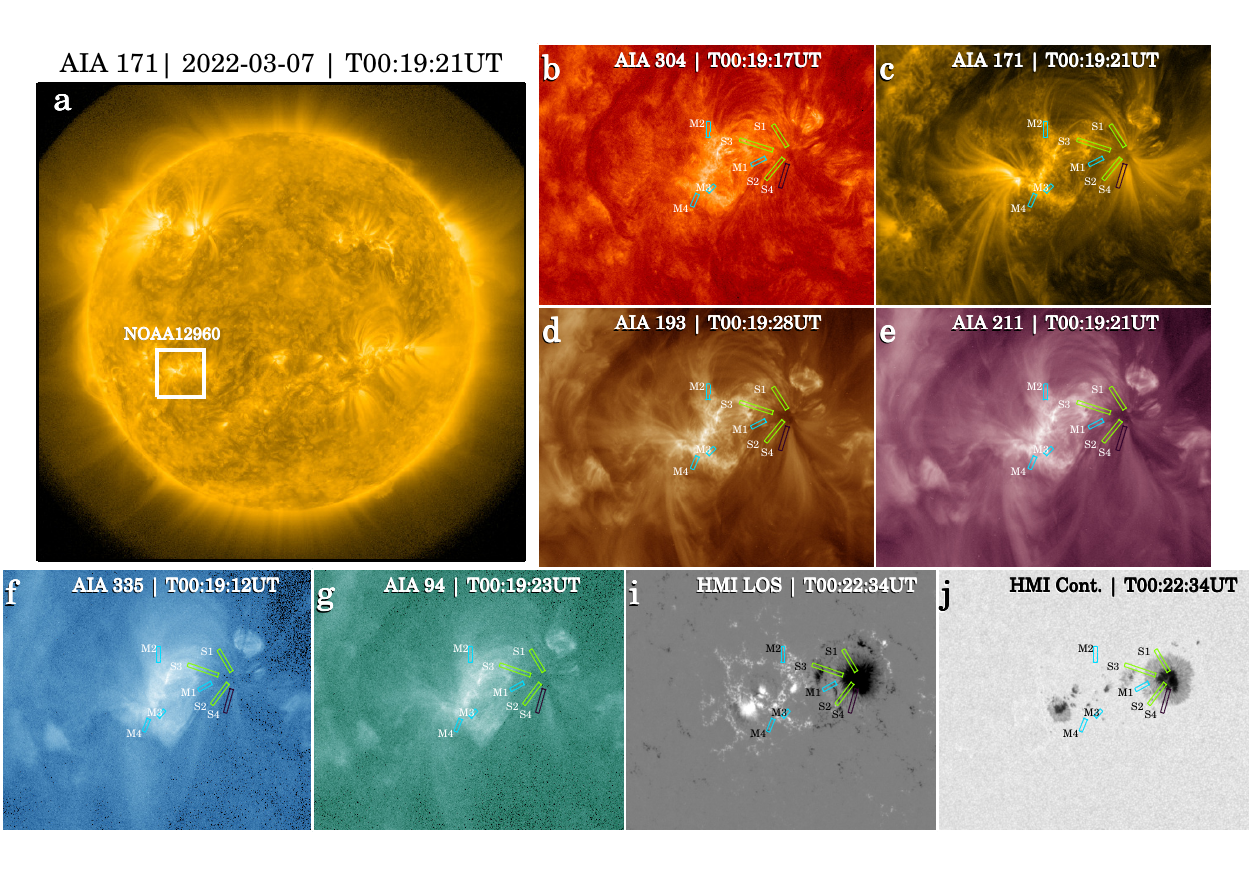}
\caption{Multi-wavelength view of the active region NOAA12960. Panel-a presents a full disc 171~{\AA} image from 2022-03-07, upon which the studied active region is outlined by the white rectangle. Zoomed in views of that rectangular region from different AIA channels are shown in panels-b-g while images from HMI LOS and continuum channels are shown in panels-i and j. Layout of the artificial slits overplotted on these panels is same as in Fig~\ref{fig:context}a. 
}
\label{fig:aia_all}
\end{figure*}

We present a multi-wavelength view of the studied active region NOAA12960 in Fig.~\ref{fig:aia_all}. Different panels in this figure display the outlook of that active region as seen through AIA and HMI channels. AIA channels such as the 94~{\AA}, 131~{\AA} and 335~{\AA} passbands, that are sensitive to emission from hotter plasma, highlight the presence of hot loops in the core of the active region. Furthermore, the footpoints of these hot loops seem to either coincide with or are located closer to the regions where M-slits are positioned. These bright footpoint regions are typically referred to as moss regions \citep{1999SoPh..190..409B}. 

\end{appendix}

\end{document}